# Joint Phase Time Array: Opportunities, Challenges, and System Design Considerations

Young-Han Nam, *Member, IEEE*, Ahmad AlAmmouri, *Member, IEEE*, Jianhua Mo, *Senior Member, IEEE* and Jianzhong Charlie Zhang, *Fellow, IEEE*

***Abstract*— This paper presents a novel approach to designing millimeter-wave (mmWave) cellular communication systems, based on joint phase time array (JPTA) radio frequency (RF) frontend architecture. JPTA architecture comprises time-delay components appended to conventional phase shifters, which offer extra degrees of freedom to be exploited for designing frequency-selective analog beams. Hence, a mmWave device equipped with JPTA can receive and transmit signals in multiple directions in a single time slot per RF chain, one direction per frequency subband, which alleviates the traditional constraint of one analog beam per transceiver chain per time slot. The utilization of subband-specific analog beams offers a new opportunity in designing mmWave systems, allowing for enhanced cell capacity and reduced pilot overhead. To understand the practical feasibility of JPTA, a few challenges and system design considerations are discussed in relation to the performance and complexity of the JPTA systems. For example, frequency-selective beam gain losses are present for the subband analog beams, e.g., up to 1 dB losses for 2 subband cases, even with the state-of-the-art JPTA delay and phase optimization methods. Despite these side effects, system-level analysis reveals that the JPTA system is capable of improving cell capacity: 5%-tile cell throughput by up to 65%. To the best of the author's knowledge, this paper is the first paper explaining the system-level benefits and system-design challenges of JPTA, with an analysis of the performance tradeoff based on an intuitive metric of beam gain losses.**

## I. INTRODUCTION

Joint-Phase Time Array (JPTA) radio frequency frontend architecture has been studied as a method for enabling frequency-selective analog beamforming, using a single or a few digital transceiver chains (TRx) [1,2]. The frequency-dependent beamforming capability has made JPTA—also referred to as true-time-delay (TTD) beamforming [3] and delay-phase precoding (DPP) [4]—a promising solution for mitigating beam squint in high-frequency bands [4], improving mmWave uplink coverage [2], enabling fast beam sweeping via rainbow beams [3,5], empowering integrated sensing and communications [6], and facilitating spectrum sharing through frequency-dependent beam nulling [7].

Although the fundamental principles of JPTA have been known for decades—particularly in phased-array radars [8]—its recent resurgence in wireless communications is driven by advances in delay-element hardware, which have made JPTA practical for mmWave networks in terms of form factor, cost, and power efficiency [9-11]. Several algorithms have been proposed to optimize JPTA parameters for linear and planar arrays [1,3,4,5,12,13,14], each balancing performance, complexity, and use-case requirements.

Traditional analog beamforming architectures impose a strict one-beam-per-TRx constraint, limiting mmWave base stations (BSs) to serving users in similar angular directions within a given scheduling slot, thereby creating scheduling bottlenecks. In contrast, JPTA enables multi-user scheduling per TRx, leveraging frequency-division multiplexing (FDM) [2] to support users in different angular directions without intra-cell interference. By relaxing scheduling constraints, JPTA has potential to enhance data throughput and reduce signaling overhead. However, practical deployment requires a careful evaluation of performance and complexity trade-offs.

Although hybrid beamforming systems with multiple TRx can also be used for generating multiple beams in the frequency domain, multiple challenges need to be resolved before they can be used in practice. Firstly, sizeable extra hardware cost and power consumption incur with equipping more TRx's, contributed by additional analog-digital and digital-analog converters (ADC/DACs) and digital modem processing requirements. Secondly, with a subarray architecture where $N$ subpanels are connected to $N$ digital chains, the beam gains per subpanel reduces by $10\log_{10}N$ from the case where a full array panel is used for the beamforming. This loss is not desirable for coverage-limited scenarios. It is also noted that the practical feasibility of fully-connected architecture to a large number of antennas (e.g., 256) has not been verified due to the wiring challenges. Owing to these challenges of multi-TRx hybrid beamforming, JPTA can be a better alternative for offering better mmWave coverage. On the other hand, JPTA has limitations over multiple TRx systems, as only multi-TRx

The authors are with Samsung Research America, Plano, TX 75024 USA. (e-mail: younghan.n, ahmad1.a, jianhua.m, jianzhong.z@samsung.com).

systems allow the possibility of allocating the same time-frequency resources to multiple users through spatial multiplexing.

This paper explores JPTA's primary use cases in commercial mmWave networks, focusing on overhead reduction and throughput enhancement. For the latter, various JPTA parameter optimization algorithms are analyzed using a beam-gain-loss metric—which directly translates to the received signal-to-noise ratio (SNR) and block error rate (BLER). While these algorithms vary in performance, the most suitable approach is not solely based on its beam gain performance, but also depends on factors such as BS architecture, real-time computational capacity, control bandwidth, and latency for setting beamforming weights in beamforming-integrated chips (BF-ICs). This paper also highlights several implementation challenges that need to be solved to facilitate adopting JPTA in a large-scale commercial setup.

To assess the ultimate system-level benefits of JPTA, a comprehensive performance analysis is conducted via simulations incorporating BS and user terminal (UT) modeling, wireless channel propagation, user scheduling, and physical-layer abstractions. Two JPTA architectures are compared—three dimensional (3D) JPTA and azimuth-only JPTA—examining their throughput gains and computational complexity to provide insights into their feasibility for next-generation mmWave networks.

The outline of this paper is as follows: In Section II, JPTA RF front-end architecture and a problem formulation to determine delay and phase parameters to construct frequency-selective analog beams will be covered. In Section III, the use cases and benefits of JPTA architecture will be introduced. In Section IV, BS system design considerations and challenges to address will be presented to realize the JPTA benefits. In Section V, a system-level JPTA performance study will be presented, demonstrating the throughput gain of JPTA against traditional single-TRx systems. There, various algorithms to derive JPTA parameters to construct desired beam patterns will be considered, and their beam gain loss performance will be presented. In Section VI, a concluding remark will be presented.

## II. Fundamentals of Joint Phase Time Array

Joint phase time array (JPTA) radio frequency (RF) frontend architecture refers to the one shown in Figure 1. The main difference between the JPTA and conventional architecture is that controllable delay elements are appended to the RF phase shifters, which offer extra degrees of freedom to design analog beams. As time-domain delay translates into frequency-domain phase ramping, the delay values assigned to the delay components can be used to create frequency selectivity for the analog beams.

Based on the equivalent frequency-domain signal model presented in Figure 1(b), the beam gain at the $k^{th}$ subcarrier can be expressed as $|\sum_{m=1}^{M} e^{j(2\pi f_k \tau_m + \phi_m)} a_m(f_k, \theta_{az}, \theta_{el})|^2$, where $a_m(\cdot,\cdot,\cdot)$ is $m^{th}$ element in the array response at subcarrier (SC)

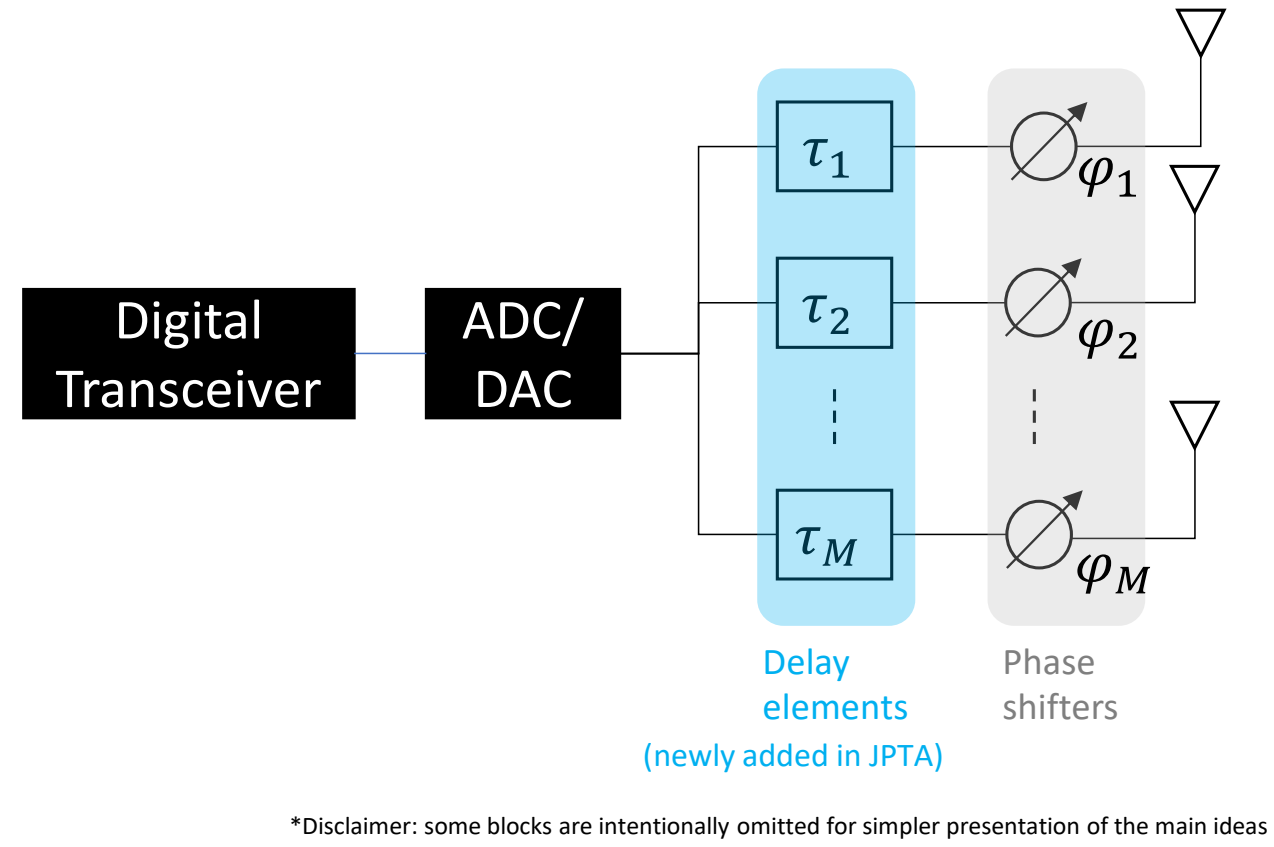

(a) JPTA RF frontend architecture

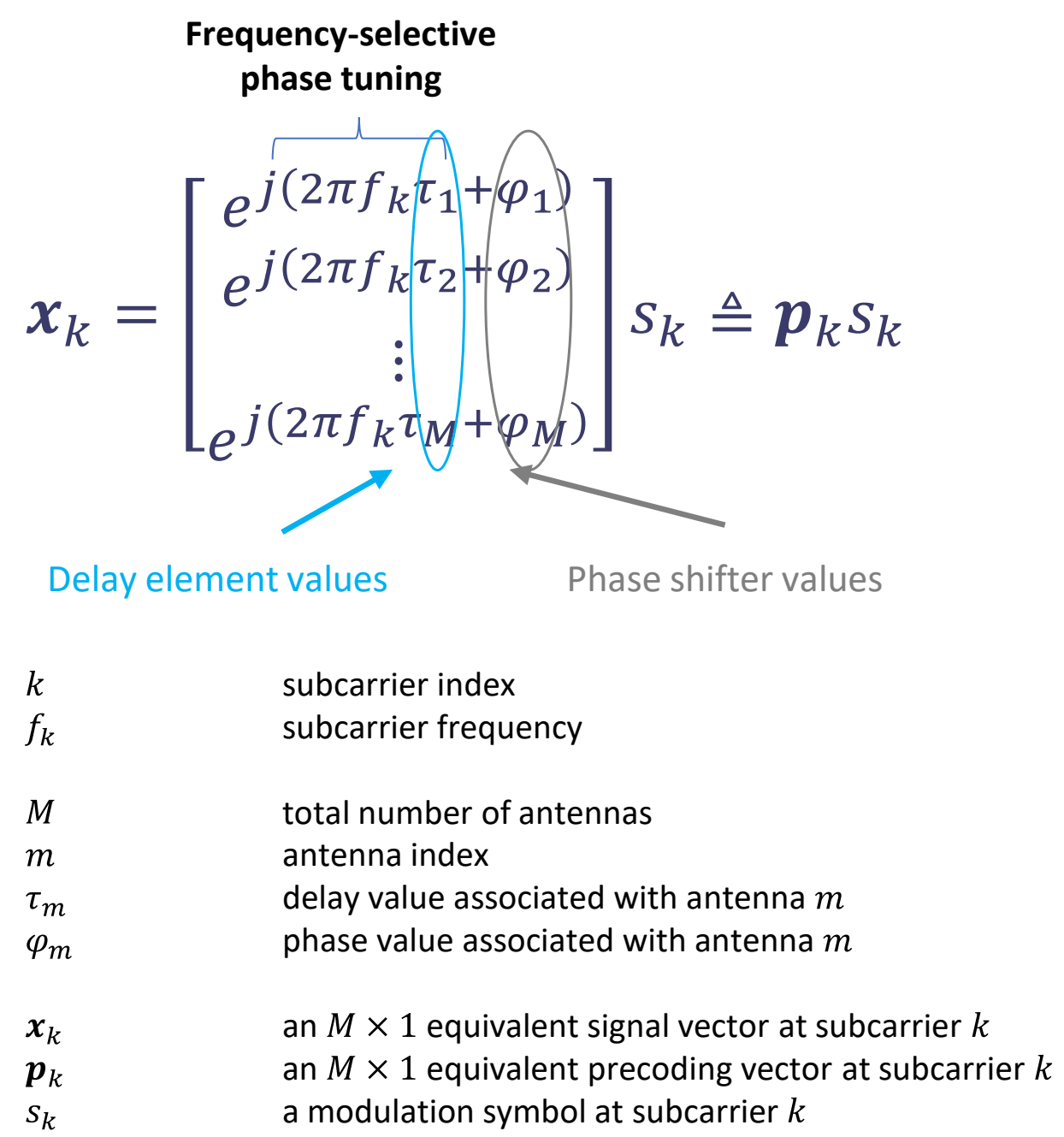

(b) Equivalent frequency-domain signal model

Figure 1. JPTA frontend architecture and an equivalent signal model in the frequency domain

frequency $f_k$, for a pair of elevation and azimuth angles of arrival (AoA): $(\theta_{el}, \theta_{az})$. For frequency-selective beamforming with different AoA pairs, subcarriers are partitioned into multiple subbands, one AoA pair per subband. The JPTA beamforming algorithm consists of finding $\tau_m, \phi_m\ \forall m \in [1, M]$ to maximize the gain over all subcarriers on their associated AoAs. For example, the mean beam gain for each AoA pair can be calculated across its sub-carriers, then the average beam gains across all of AoAs can be the objective function for the optimization problem.

## III. Use Cases and System Benefits

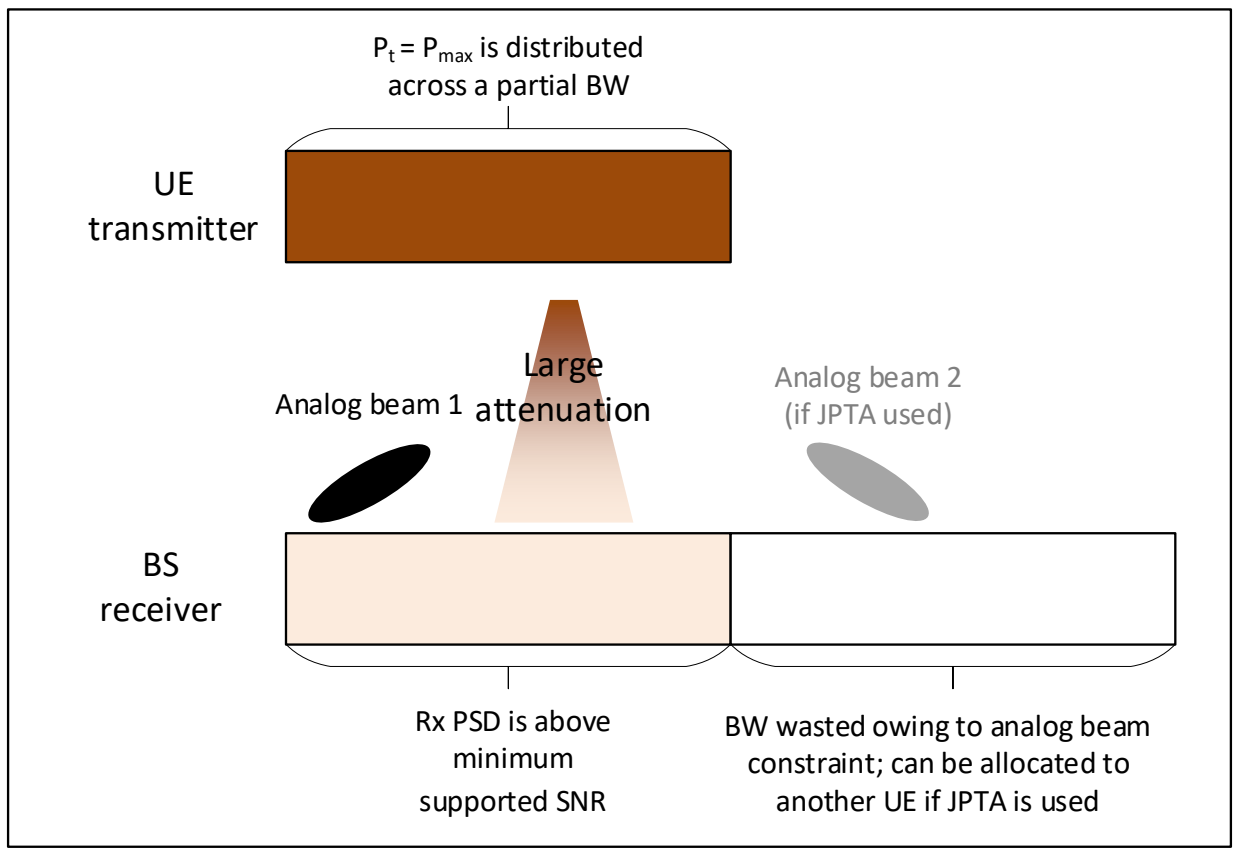

(a) For FDM of power-limited UEs' transmissions

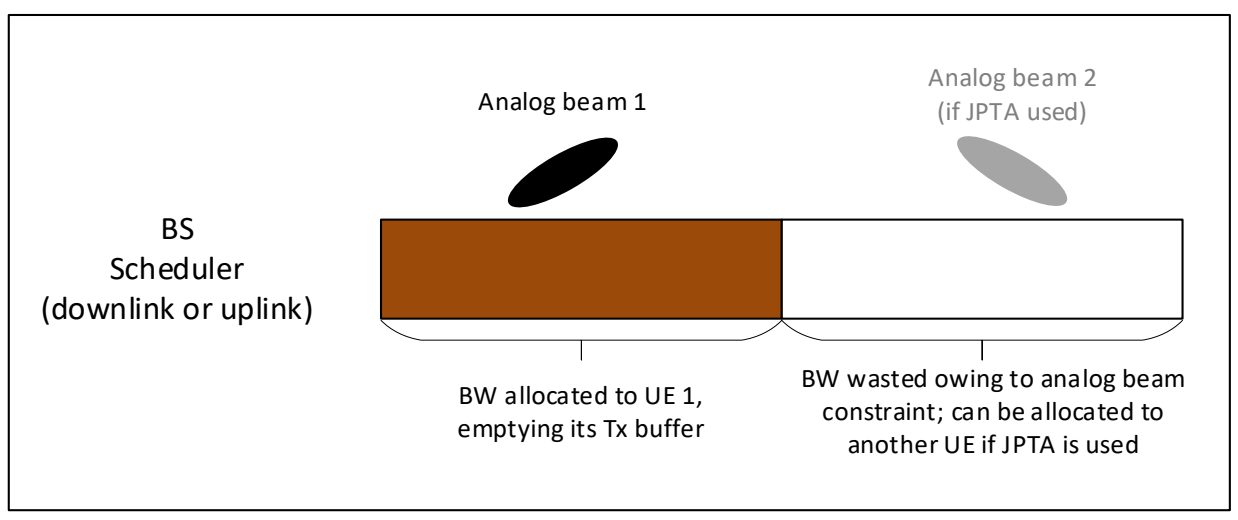

(b) For partially-loaded UEs

Figure 2. JPTA use cases for data channels

By exploiting frequency-selective analog beams constructed with JPTA, a base station can be designed to simultaneously serve multiple users in different angular locations in a frequency-division multiplexing (FDM) manner, with a single digital transceiver chain.

In this sequel, we will discuss system design possibilities with JPTA and the potential benefits of these designs.

### *A. Data Channels*

As JPTA does not support multi-user multi-input-multi-output (MIMO) spatial multiplexing transmissions, the peak spectral efficiency of a JPTA system remains the same as that of a conventional single-TRx system. However, JPTA helps to achieve extra cell capacity in certain cases, namely for power-limited uplink transmissions and for partially loaded UTs as illustrated in Fig. 2.

The use case for power-limited UTs studied is shown in Fig. 2(a). To comply with industry standards and meet regulatory requirements, UT's maximum transmission power shall not exceed a certain value, say $P_{max}$. Suppose that a power-limited UT is configured to transmit with $P_{max}$. In such a case, the serving BS typically also limits the UT's transmission BW to ensure successful demodulation of a transmitted data channel with a minimum required SNR despite the expected channel attenuation. In other words, the power spectral efficiency (PSD) has to be above a certain threshold for successful transmission.

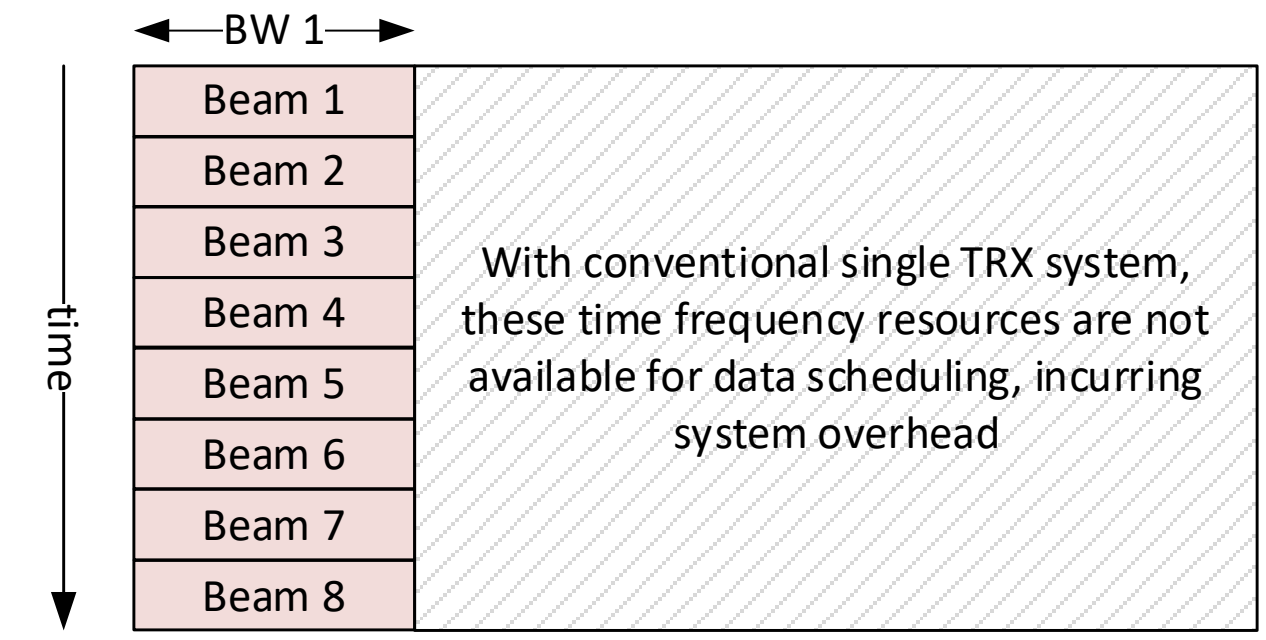

(a) In conventional single-TRx systems

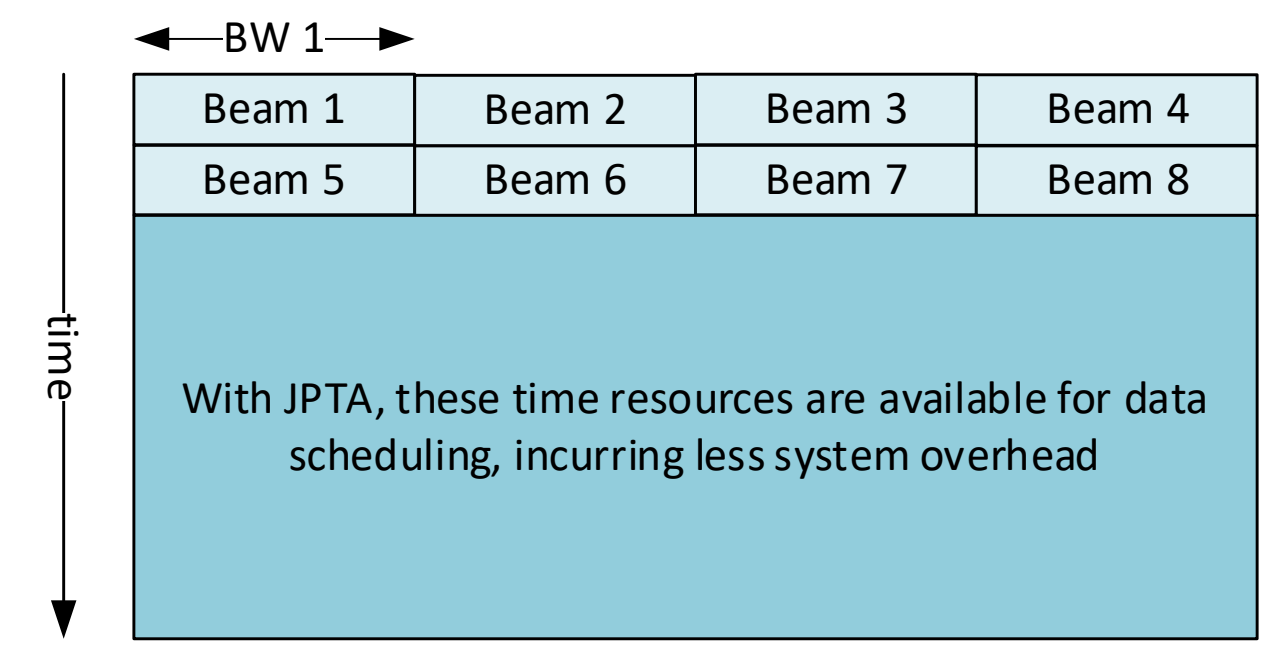

(b) In systems with JPTA

Figure 3. Multiplexing beams in conventional single TRx systems vs. JPTA systems

When the UT's scheduled transmission BW is smaller than the system bandwidth and conventional analog beam constraint is imposed, the left-over bandwidth is wasted if no users are available in the same angular direction as the power-limited UT. With JPTA, the BS can allocate the remaining bandwidth to other UTs in different angular directions, which leads to improved cell throughput. Also, as power-limited UTs are scheduled more frequently with JPTA, a UT's UL throughput at a given location increases, or a target UL data rate can be achieved at a farther distance. Exploiting this, a new scheduler can be devised to maximize scheduling opportunities for individual users while allowing a reduction of per-user scheduling bandwidth. With the new scheduler, the overall user throughput increases further thanks to more scheduling opportunities and higher per-tone transmission power [2].

Fig. 2 (b) presents the use case for partially loaded UTs. With partial loading, the amount of data to be transmitted to or received by a UT is finite, sometimes insufficient to fill up the entire system BW. With JPTA, the left-over bandwidth can be scheduled for another UT in different angular locations, which helps to improve cell throughput and user perceived throughput.

### *B. Pilots and Overhead Channels*

Frequency-selective analog beams generated with JPTA can be used for reducing overhead by multiplexing beams in the same time duration for beam management pilots and control channels [3-5]. In conventional single TRx systems, these overhead channels are transmitted in such a way that a single beam is transmitted per time duration. When these overhead channels do not require full bandwidth usage for the scheduled time duration, the remaining bandwidth is wasted, as illustrated in Fig. 3(a). With JPTA, on the other hand, these overhead channels intended for users on different angular locations are scheduled with these frequency selective analog beams in the same time duration, as presented in Fig. 3(b). Accordingly, more full-bandwidth time resources are open for data transmissions, which can be used to increase cell throughput.

## IV. System Design Challenges and Considerations

Prior work has demonstrated the implementation of efficient CMOS-based delay elements with a wide delay range [9] and the feasibility of JPTA integration in a proof-of-concept (PoC) setup [10]. Advancing from the prior work, this section discusses challenges associated with large-scale commercial deployment of JPTA. Specifically, key aspects such as channel estimation, codebook design, hardware simplifications, and scheduling are discussed.

**Channel estimation**: To perform FDM of data channels with different beams as explained in Section III-A, the BS needs to know the best beams for different UTs and other conventional scheduling parameters for the UTs, e.g., buffer status, channel condition, etc. Such information is readily available in single TRx conventional systems, relying on beam management and channel state information (CSI) feedback constrained on single TRx operations as defined in 3GPP [15]. For reducing initial implementation complexity, one possible option for the JPTA BS's acquiring of the necessary beam and scheduling information is to turn JPTA off for CSI acquisition. This requires that JPTA delay elements should be able to be dynamically configured to zero as well as non-zero values.

**Scheduling**: To schedule a JPTA data channel for a time slot, multiple pairs of a scheduling subband (SB) and a beam angle need to be determined to serve multiple UTs. Having individual UTs' scheduling parameters, the BS can rely on traditional scheduling metrics, e.g., proportional fair metrics, to choose a primary user to schedule. The rest of the bandwidth can be allocated to other UTs even if they are on different beam directions, considering tradeoffs among cell capacity, fairness, priority, etc. To further improve user throughput, the BS scheduler may want to reduce the primary user's scheduling bandwidth with the intention of increasing per-user scheduling opportunities [2].

**Codebook design**: For online beamforming, the beamforming algorithm and configuration of the BFICs must be fast to accommodate the short time slot in mmWave bands. Alternatively, an offline beamforming scheme can be adopted, where a codebook mapping beam IDs to phase and delay values is designed offline and stored at the BFICs. However, this introduces limitations on the codebook size due to the limited memory storage at the BFICs. Hence, whether an offline or an online approach is adopted, innovation is needed to facilitate integrating JPTA in a large-scale commercial setup.

**Standards protocols**: In the current 3rd generation partnership project (3GPP) New Radio (NR) specifications related to mmWave beam management and data channel transmissions and receptions, there is an implicit assumption of analog beamforming constraints. Full benefits of JPTA can be exploited if the specifications are designed with considering JPTA capability. Pilot channels such as synchronization signal blocks (SSB), channel-state-information reference signals (CSI-RS), etc., could be re-designed with exploiting JPTA for enabling FDM beam sweeping. Control channels such as physical uplink control channels (PUCCH) and physical downlink control channels (PDCCH) could also be re-designed.

**Hardware simplification**: For reducing hardware design costs, spatial-domain simplification can also be considered. Instead of having one delay component per antenna to make the JPTA beams distributed in both azimuth and elevation directions, the JPTA system can be constructed with one delay component for each column of the uniform rectangular antenna array to make JPTA beams distributed across the azimuth domain only. This incurs extra scheduling constraints, and the performance impact of this hardware simplification will be analyzed in Section V-C.

## V. Numerical Performance Analysis

This section presents a numerical performance analysis that reveals the practical performance benefits of JPTA compared to baseline phased antenna array systems.

To facilitate this comparison, we introduce an intuitive metric, ***average beam gain loss***, for evaluating different algorithms used to derive the JPTA parameters. Frequency-selective analog beams generated by JPTA typically suffer reduced beam gains within the target subbands, impacting performance. Since decoding performance in frequency-selective fading channels correlates directly with the average SNR across frequency subbands [2], and average SNR (in dB) is linearly related to average beam gain loss (in dB), a smaller beam gain loss translates directly to a lower block error rate.

The analysis is on the JPTA use case of uplink throughput enhancement, particularly for cell-edge users.

### *A. System Model*

A single-TRx BS equipped with a 24V16H antenna array (24 vertical and 16 horizontal antenna elements) is assumed, with each antenna element (AE) controlled by a 6-bit tunable phase shifter. The carrier frequency is set to 28GHz with a bandwidth of $B_{total}$=400 MHz and a SC spacing of 120 kHz. The BS's horizontal scan range extends 120°, and the vertical scan range covers 25°, divided uniformly across 126 beams (18 rows and 7 columns).

Two JPTA architectures are analyzed: 3D JPTA and azimuth-only (AO) JPTA. In 3D JPTA [12], each AE is paired with an adjustable delay element, enabling frequency-dependent beamforming across both azimuth and elevation planes. In contrast, AO JPTA connects each column of antennas to a single delay element—yielding a total of 16 delay elements—allowing frequency-dependent beamforming only in the azimuth plane, with elevation control handled by phase shifters. While 3D JPTA offers increased beamforming flexibility, it requires more delay elements to achieve its functionality.

### B. Beam-Gain Loss

JPTA beam design algorithms vary in terms of performance and complexity. Options include a closed-form least squares (LS) solution in [14] (extended to 3D in [12]) for low complexity, an iterative algorithm [1] offering a balanced performance-complexity trade-off, and a gradient descent (GD) approach [12] providing high beam gains at the cost of increased computational complexity. The choice of beam design algorithm depends on the BS's computational resources and whether online or offline beam design is feasible. Effective beam gain for JPTA is computed by averaging dB beam gains across SCs assigned to each UT. To fairly assess the performance of the different algorithms, we rely on the mean beam gain loss defined as

$$L = 10\log_{10} N_t - \frac{1}{N_{UE}}\sum_{i=1}^{N_{UE}} \bar{G}^{i}{}_{dB}\ ,$$

where $N_t$ is the number of transmit antennas and $\bar{G}^{i}{}_{dB}$ is the beam gain of the $i^{th}$ UT averaged in dB over the subcarriers assigned to it.

Table I presents beam gain losses (in dB) evaluated for different algorithms and beam configurations based on simulations of 100,000 UTs uniformly distributed in the angular domain. The GD algorithm, as expected, achieves the lowest beam gain loss—up to 1.43 dB (0.3 dB) less at the 90th percentile relative to LS (iterative) algorithm—though it demands higher computational power. The iterative algorithm achieves up to 1.2 dB improvement over LS, balancing performance with lower complexity. In particular, the 90%-tile beam gain loss remains below the 3 dB and 6 dB bounds for two-beam and four-beam configurations, respectively. Here, 3 dB and 6 dB bounds represent the beam gain losses from constructing 2 and 4 beams with a single set of phase shifters without introducing JPTA. Additionally, Table I highlights the maximum delay required to achieve the desired beam patterns, observed below 10 ns, which sets a requirement for the delay component design to construct up to 4-beam JPTA systems for 400 MHz operational bandwidth.

Table I: JPTA beam gain loss in dB

| JPTA Type | # of beams | Algorithm | 90%-tile loss (dB) | Mean loss (dB) | Max delay (ns) |
|---|---|---|---|---|---|
| AO | 2 | LS | 1.04 | 0.84 | 3.5 |
| | | Iterative | 1.03 | 0.80 | 3.5 |
| | | GD | 1.00 | 0.79 | 3.5 |
| | 4 | LS | 4.65 | 2.67 | 9 |
| | | Iterative | 3.41 | 2.03 | 9 |
| | | GD | 3.22 | 2.00 | 9 |
| 3D | 2 | LS | 0.99 | 0.98 | 3.5 |
| | | Iterative | 1 | 0.95 | 4 |
| | | GD | 0.99 | 0.94 | 4 |
| | 4 | LS | 4.56 | 3.34 | 9.5 |
| | | Iterative | 3.5 | 2.56 | 9.5 |
| | | GD | 3.2 | 2.53 | 9.5 |

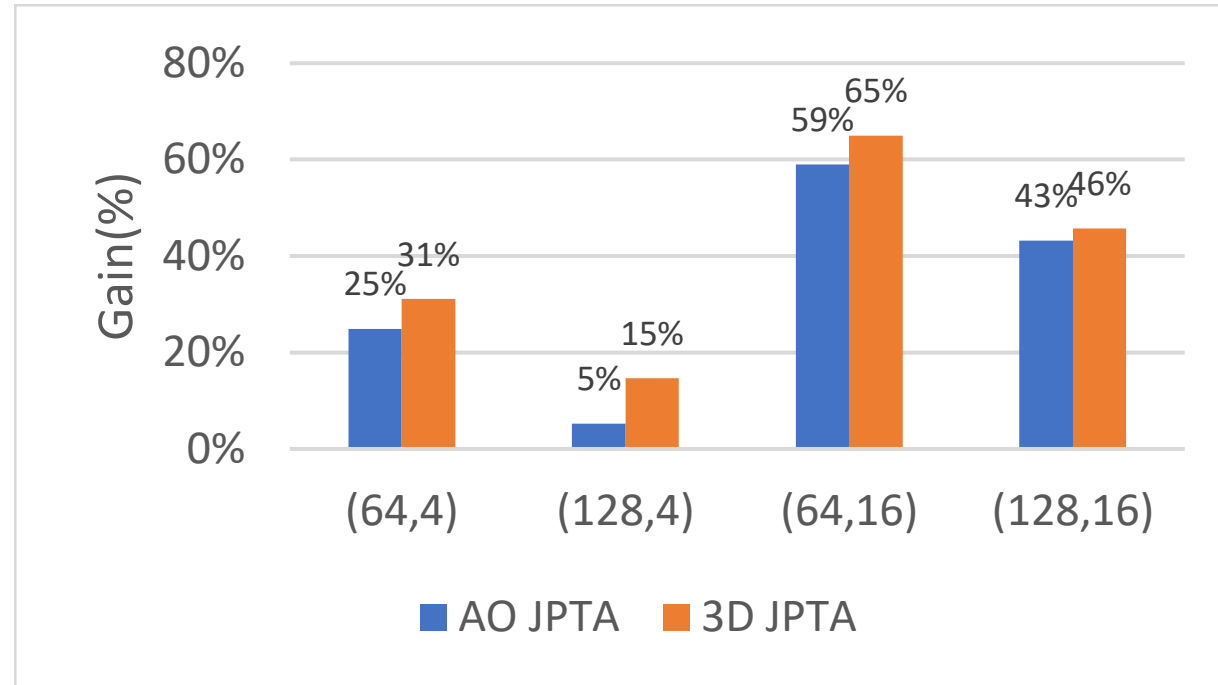

(a) 5%-tile

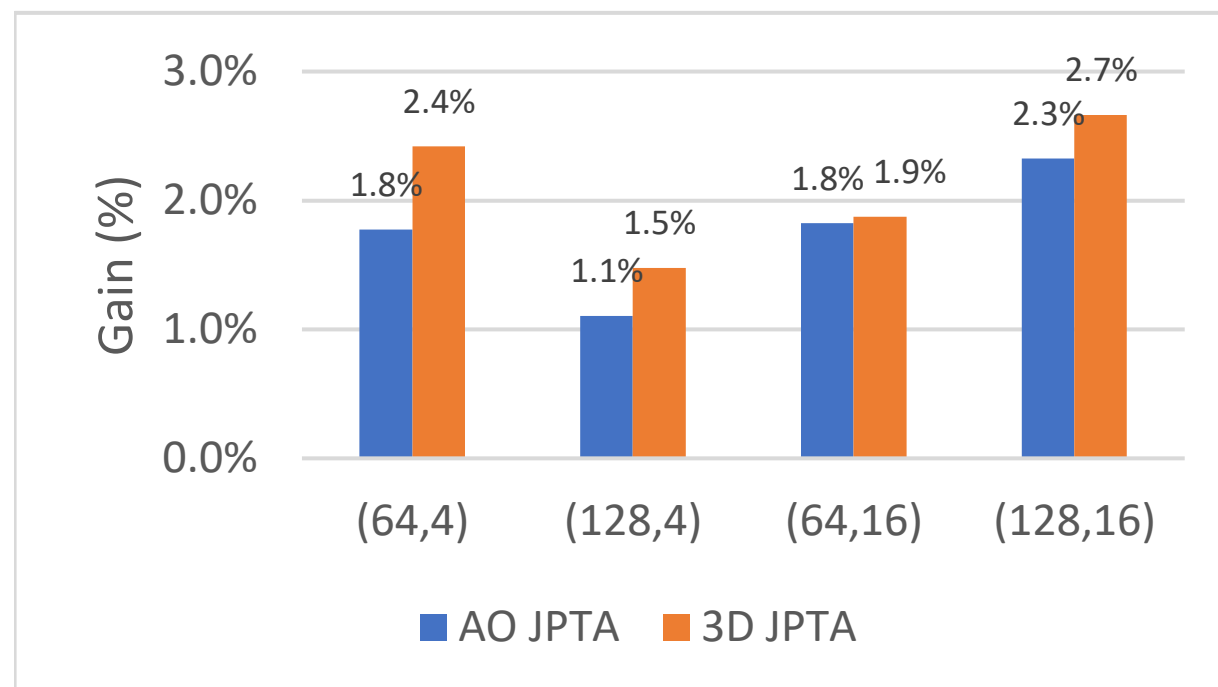

(b) Average

Figure 4. Throughput gain of AO and 3D JPTA for different $(N_{pool}, N_{max})$ combinations

Another insight from Table I is the beam gain loss of 3D vs AO. The flexibility of 3D JPTA in pairing UTs across different elevation angles comes at a cost of higher mean beam gain loss. Note that although we focused on the case of 120 kHz sub-carrier spacing, we observed that the beam gain loss is effectively the same for spacings ranging from 15 kHz to 480 kHz.

### C. System Throughput Analysis

A system-level simulation is performed to evaluate JPTA's impact on throughput. The study area is a 3 km radius around a site in Flower Mound, TX, USA, encompassing 5,000 houses within range, and wireless channels of BS-UT links were generated by ray-tracing simulations. Fixed wireless access (FWA) scenario is evaluated, where each house features wall-mounted customer premises equipment (CPE), i.e., a type of UT used for FWA, with a single-TRx 8V8H array. These CPEs are aligned toward the strongest BS path, achieving an effective isotropic radiated power (EIRP) of 51 dBm. For each evaluation, out of the 5000 CPEs, a pool of $N_{pool}$ CPEs is randomly selected. For simplicity, round-robin scheduling of up to $N_{max}$ CPEs is conducted per scheduling time slot to model BS scheduling and beamforming constraints.

For the baseline, UTs are scheduled together only when the same beam serves them. For AO JPTA, UTs are scheduled together if the same elevation beam serves them, while in 3D JPTA, UTs are scheduled together without restrictions on their serving beams. In JPTA, scheduling priority for secondary UTs is first given to those served by the same beam as the primary UT, then to UTs on the same elevation beam row, and then for 3D JPTA, the rest of the UTs. This priority hierarchy minimizes beam gain loss and activates JPTA selectively.

To intuitively understand the impact of JPTA on UT throughput, power-limited UTs with low SNRs scheduled over a narrow bandwidth $B < \frac{B_{total}}{N_{beams}}$ are considered, where $N_{beams}$ represents the number of JPTA beams (which will be UT multiplexing gain). For these UTs, throughput $R$ is approximated using Shannon's formula for low SNR as $R \approx B\,T\,\mathrm{SNR}$, where $T$ is the time fraction allocated to the UT without JPTA and SNR assumes the full beamforming gain of $10\log_{10} N_t$. With JPTA, the total throughput becomes $R_{JPTA} \approx \frac{B\,T\,N_{beams}\,SNR}{l(N_{beams})}$, where $l(N_{beams})$ accounts for the beam gain loss. This implies that JPTA enhances performance for power-limited UTs as long as $\frac{N_{beams}}{l(N_{beams})} > 1$. For example, in two-beam (four-beam) scenarios, beam gain loss must remain below 3 dB (6 dB) for JPTA to be beneficial.

To understand the impact of JPTA on all UTs, Figure 4 presents mean and 5%-tile throughput gains for different $(N_{pool}, N_{max})$ combinations, relative to traditional analog beamforming, with JPTA beam count capped at four to limit the codebook or beamforming design complexity. As the figure shows, JPTA delivers improvements in average throughput and 5%-tile throughput (representing cell-edge UTs). For fixed $N_{max}$, increasing $N_{pool}$ generally reduces JPTA throughput gains, as more UTs may share the same beam, reducing the need for JPTA. In the AO-JPTA example, the 5%-tile throughput gain drops from 25% to 5% when the UT density doubles with a fixed $N_{max} = 4$. Conversely, increasing $N_{max}$ for a fixed $N_{pool}$ enhances JPTA throughput gains, as fewer UTs will likely be served by the same beam. In the AO-JPTA example, the 5%-tile throughput gain increases from 5% to 43% when $N_{max}$ increases from 4 to 16 for a fixed UT density. Thus, JPTA performance improves when the BS can schedule more UTs per slot, supporting the use of low-complexity scheduling and beamforming algorithms.

High-SNR users achieve limited benefits from JPTA, as they are not power-limited and can utilize increased transmission opportunities in the time domain as effectively as additional bandwidth. Conversely, cell-edge UTs, constrained by instantaneous power limitations, benefit from increased transmission time, even if it comes with reduced bandwidth. This enhancement for cell-edge UTs effectively expands broadband coverage, addressing the common constraint of uplink throughput in FWA scenarios. These trends hold consistently across both AO and 3D JPTA configurations, highlighting JPTA's impact on improving uplink performance, particularly for cell-edge users. Compared to AO JPTA, 3D JPTA achieves higher throughput gains due to its higher scheduling flexibility, despite having higher beam gain loss in some cases. With simple calculations in Figure 4, 3D JPTA provides 2.1%-9.5% gain relative to the AO JPTA in terms of the 5%-tile throughput, i.e., cell-edge UTs. The gain is more limited for the mean throughput, since it is below 1% in the considered cases. Overall, 3D JPTA's requirement for more delay elements introduces a performance-cost trade-off, balancing enhanced throughput with added hardware complexity.

## VI. Concluding Remarks

This paper introduced system benefits, system design considerations, and system performance of mmWave base station (BS) systems equipped with Joint Phase Time Array (JPTA) radio-frequency (RF) frontend architecture. With JPTA, BS is capable of scheduling multiple devices in different spatial directions in the same time slot, alleviating traditional analog beamforming constraints. The frequency-selective beams formed by the JPTA systems were proposed to be used to improve data channel coverage and capacity, and to reduce overhead and latency for beam management. Then practical system design constraints were presented, and various options to reduce system design complexity were discussed, e.g., azimuth-only JPTA instead of three dimensional JPTA. System-level simulation results confirmed that JPTA improves average and cell-edge throughput, with the most substantial gains observed at the cell edge (up to 65% throughput gain), often constrained in fixed-wireless-access settings. The findings showed that high-SNR users see minimal impact, while cell-edge users benefit more from JPTA due to the increased transmission time.

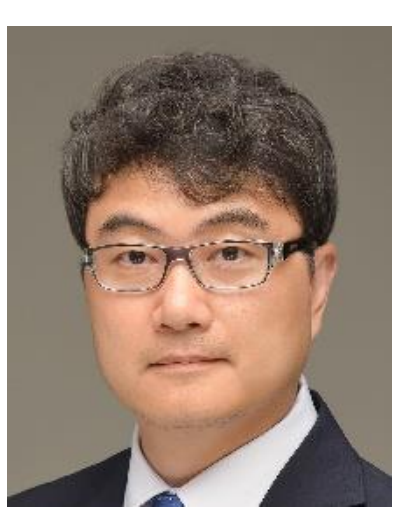

**Young-Han Nam** (Member, IEEE) received Ph.D. from the Ohio State University in 2008, and MS and BS degrees from Seoul National University in 2002 and 1998. From 2008 to 2019, he was with Samsung Research America (SRA), making contributions to the 3GPP RAN1 standards on mmWave, massive MIMO, 5G channel modeling, etc. For 3GPP standardization, he served as a rapporteur for the 5G channel modeling study item, responsible for editing TR38.901. From 2019 to 2022, he was with Mavenir, overseeing O-RAN product R&D in the system engineering team. Currently, he serves as a Senior Director in SRA, leading 6G cellular communication system research.

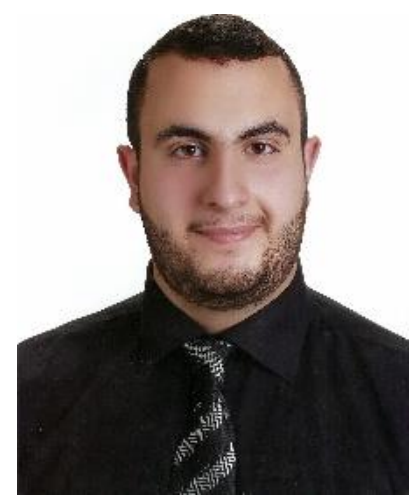

**Ahmad AlAmmouri** (Member, IEEE) received the B.Sc. degree (Hons.) from the University of Jordan, Amman, Jordan, in 2014, the M.Sc. degree from the King Abdullah University of Science and Technology (KAUST), Thuwal, Saudi Arabia, in 2016, and the Ph.D. degree from The University of Texas at Austin, Austin, TX, USA, in 2020, all in electrical engineering. He is currently a Staff Research Engineer with Samsung Research America, Plano, TX, USA. He was awarded the Chateaubriand Fellowship from the French Embassy in the USA, in 2019, the Professional Development Award from UT Austin, in 2019, and the WNCG Student Leadership Award, in 2020. He was recognized as an Exemplary Reviewer from the IEEE TRANSACTIONS ON COMMUNICATIONS, in 2017, and from IEEE TRANSACTIONS ON WIRELESS COMMUNICATIONS, in 2017 and 2018.

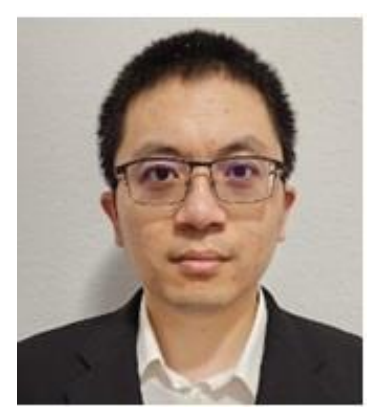

**Jianhua Mo** (Senior Member, IEEE) received the B.S. and M.S. degrees from Shanghai Jiao Tong University in 2010 and 2013, respectively, and the Ph.D. degree from The University of Texas at Austin in 2017, all in electronic engineering. He is a Senior Staff Engineer with Samsung Research America, Plano, TX, USA. His research interests include physical layer security, MIMO communications with low-resolution ADCs, and mmWave beam codebook design and beam management. His awards and honors include the Heinrich Hertz Award in 2013, the Stephen O. Rice Prize in 2019, the Best Wi-Fi Innovation Award by Wireless Broadband Alliance (WBA) in 2019, an Exemplary Reviewer of the IEEE WIRELESS COMMUNICATIONS LETTERS in 2012, an Exemplary Reviewer of the IEEE COMMUNICATIONS LETTERS, in 2015, and the Finalist for Qualcomm Innovation Fellowship, in 2014.

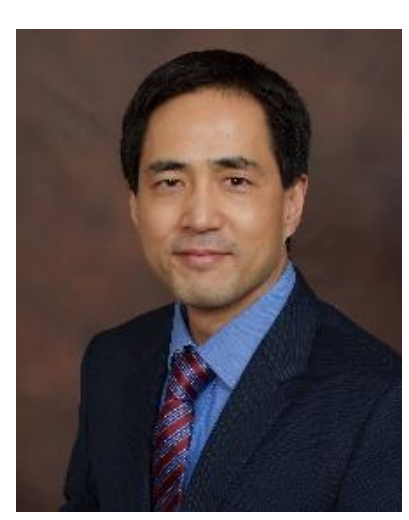

**Jianzhong Charlie Zhang (**Fellow, IEEE) received the Ph.D. degree from the University of Wisconsin-Madison. He was with the Nokia Research Center, from 2001 to 2006, and Motorola, from 2006 to 2007. From 2009 to 2013, he was the Vice Chairman of 3GPP RAN1 WG. He is currently the Senior Vice President and the Head of the Standards and Mobility Innovation Laboratory, Samsung Research America, where he leads research, prototyping, and standards for 5G cellular systems and future multimedia networks.